\documentclass[prd, aps, superscriptaddress, preprintnumbers, twocolumn, floatfix, nofootinbib]{revtex4}

\usepackage{amsfonts}
\usepackage{amsmath}
\usepackage{amssymb}
\usepackage{bm}
\usepackage{dcolumn}
\usepackage{epsfig}
\usepackage{graphicx}   
\usepackage{graphics}
\usepackage[latin1]{inputenc}
\usepackage{latexsym}
\usepackage{rotating}
\usepackage{hyperref}

\usepackage{amsfonts}
\usepackage{amsmath}
\usepackage{amssymb}
\usepackage{bm}
\usepackage{dcolumn}
\usepackage{epsfig}
\usepackage{graphicx}
\usepackage{graphics}
\usepackage[latin1]{inputenc}
\usepackage{latexsym}
\usepackage{rotating}
\usepackage{hyperref}

\newcommand\be{\begin{equation}}
\newcommand\ba{\begin{eqnarray}}
\newcommand\ee{\end{equation}}
\newcommand\ea{\end{eqnarray}}

\begin{document}

\title{Particle Production in Ekpyrotic Scenarios}

\author{W. S. Hip\'olito-Ricaldi}
\email{wiliam.ricaldi@ufes.br}
\affiliation{Physics Department, McGill University, Montreal, QC, H3A 2T8, Canada, and\\
 Departamento de Ci\^encias Naturais, Universidade Federal do Esp\'{\i}rito Santo,
Rodovia BR 101 Norte, km. 60, S\~ao Mateus, ES, Brazil}

\author{Robert Brandenberger}
\email{rhb@physics.mcgill.ca}
\affiliation{Physics Department, McGill University, Montreal, QC, H3A 2T8, Canada, and \\Institute for Theoretical Studies,
ETH Z\"urich, CH-8092 Z\"urich, Switzerland}

\author{Elisa G.M. Ferreira, L. L. Graef}
\email{@physics.mcgill.ca}
\affiliation{Physics Department, McGill University, Montreal, QC, H3A 2T8, Canada}

\date{\today}

\begin{abstract}

We consider Parker particle production in the Ekpyrotic scenario (in particular in the
New Ekpyrotic model) and show that
the density of particles produced by the end of the phase of Ekpyrotic contraction is
sufficient to lead to a hot state of matter after the bounce. Hence, no separate reheating
mechanism is necessary.

\end{abstract}

\maketitle

\section{Introduction} 

There is overwhelming observational evidence that
there was a early phase in the evolution of the universe
when Standard Model matter was in thermal equilibrium (see
e.g. \cite{Mukhanov, Weinberg, Peter} for modern textbooks
on cosmology). The best quantitative evidence for such
an early phase comes from the black body nature of the
cosmic microwave background \cite{Halpern, COBE} and
from the abundances of light nuclei (see e.g. \cite{Weinberg2}).

In Standard Big Bang cosmology it is assumed that the Universe
begins in a hot and dense thermal state. However, this scenario
cannot explain the isotropy of the cosmic microwave background,
and it cannot explain the origin of observed inhomogeneities on
length scales which were larger than the Hubble radius at the
time $t_{eq}$ of equal matter and radiation.

The inflationary universe scenario \cite{Guth} (see also \cite{Brout, Sato, Fang, Kazanas, Starob1}) 
provides a solution to these problems of Standard Big Bang cosmology,
and at the same time yields a causal theory for the origin of the
inhomogeneities which are now explored through cosmological
observations \cite{Mukh} (see also \cite{Press}). The inflationary
scenario posits a phase of almost exponential expansion of the
early universe. This phase leaves behind a vacuum state of the
Standard Model matter fields. Hence, in order to make contact with
the late time cosmology, a new phase must be posited during which
Standard Model matter fields approach a hot thermal state. This
phase is called {\it reheating phase}, and is an essential part of
the inflationary scenario (see e.g. \cite{DL, AFW} for initial studies
of this phase). Reheating in the inflationary scenario is a
semiclassical effect. It relies on the squeezing of quantum
vacuum perturbations left behind at the end of the inflationary
phase. It is a consequence of coupling terms between the scalar field
$\phi$ which generated inflationary expansion and the Standard
Model fields. As first studied in \cite{DK, TB} and worked out in more
detail in \cite{STB, KLS2} (see \cite{RehRevs} for recent reviews
on reheating), the energy transfer from $\phi$ to Standard Model fields
proceeds via a parametric instability in the equation of motion for
the Standard Model fields induced by the time dependence of $\phi$.
As a consequence, it is expected that most of the energy density $\rho({t_R})$
at the final time $t_R$ of the phase of exponential expansion ends
up in the hot plasma of the post-inflationary universe \footnote{Note that
the state of Standard Matter fields after preheating does not have a thermal
distribution of momenta.}. 

Inflation is not the only early universe scenario which is compatible
with cosmological observations (see e.g. \cite{RHBrevs} for recent
reviews of alternatives). Bouncing cosmologies with an initial
matter-dominated phase of contraction (see \cite{MBrev} for a recent
review of this {\it matter bounce} scenario) 
also produce a spectrum of fluctuations compatible with
observations, as does {\it string gas cosmology} \cite{BV} (see also
\cite{SGCrevs} for reviews) and the {\it Galileon genesis
scenario} \cite{Galileon}, two early universe scenarios of
emergent type. However, as emphasized in \cite{Peter2}, bouncing
cosmologies suffer from an instability of the contracting phase to the
growth of anisotropies. The Ekpyrotic scenario \cite{Ekp} is
an alternative to cosmological inflation which is free of the
anisotropy problem \cite{NoBKL}. 

It is of interest to study how the late time hot thermal state emerges in
models alternative to inflation, and whether it is necessary to
introduce a new mechanism analogous to inflationary reheating.
In {\it string gas cosmology} the hot thermal state of matter emerges
directly after the initial stringy Hagedorn phase \cite{BV}, and no
additional physics must be added to the system. Defrosting at
the end of the Galileon phase was studied in \cite{defrost}, and
was shown to rely on analogous coupling terms as are used in
inflationary cosmology. On the other hand, it was shown in \cite{Jerome}
that Parker particle production \cite{Parker} is sufficient to produce
a hot early universe in the {\it matter bounce} scenario. Here, we
study matter particle production in the {\it New Ekpyrotic} scenario \cite{new}
(see also \cite{new2}),
a version of the Ekpyrotic scenario which generates an approximately
scale-invariant spectrum of cosmological perturbations. We show
that the analysis of \cite{Jerome} carries over and that Parker particle
production is sufficient to lead to a hot thermal state of matter \footnote{As
in the case of preheating after inflation, the state which results from
Parker particle production does not have a thermal distribution of
excitations, but it is a state which consists of particle excitations
as opposed to being a coherent homogeneous condensate, a state which
can subsequently thermalize via particle interactions. Whereas
preheating after inflation produced mainly infrared excitations,
we will see that Parker particle production in the New Ekpyrotic
scenario leads to a spectrum of particles peaked in the ultraviolet,
and hence closer to a thermal equilibrium state.}  
 
In the following section we briefly review Parker particle production,
and in the third section we summarize the application to the
New Ekpyrotic scenario. To set our notation, we use units
in which $c = k_b = 1$, and the usual space-time coordinates
in which the background metric is
\be
ds^2 \, = \, dt^2 - a(t)^2 dx^2 \, ,
\ee
where $t$ is physical time and $x$ are the comoving spatial
coordinates (we take the spatial sections to be flat). Linear
metric and matter fluctuations about this background have
a complete basis of fundamental solutions which are
Fourier modes in comoving spatial coordinates with a
time-dependence which depends on the background.
The comoving momentum is denoted by $k$. An
important length scale is the Hubble radius
\be
l_H(t) \, \equiv \, H(t)^{-1} \, \equiv \, \frac{a(t)}{\dot{a}(t)} \, .
\ee
The Hubble radius separates length scales where
the mode evolution is qualitatively different.
               
\section{Review of Parker Particle Production} 

Parker particle production \cite{Parker} (see also \cite{Mosto}) is
a phenomenon which was first studied in the context of quantum
field theory in curved space-time. If we assume that the space-time
is Minkowski space-time both at early and at late times, but underwent
a period of expansion during an intermediate time interval, then the
initial Minkowski vacuum state of a test scalar field $\chi$
will evolve non-trivially during the intermediate time interval, and 
evolve into a final state of $\chi$ which is not equal to the final time Minkowski 
vacuum. From the point of view of the final Minkowski frame, the 
final state contains $\chi$ particles. 

The same phenomenon applies to linear cosmological perturbations
which evolve in a similar way to test scalar fields on the cosmological 
background (see e.g. \cite{MFB, RHBfluctrev} for reviews of the theory 
of cosmological perturbations). On length scales smaller than the
Hubble radius $H^{-1}(t)$, where $H(t)$ is the expansion rate,
fluctuations oscillate as they do in Minkowski space-time, whereas
they are squeezed on super-Hubble scales. The squeezing of
fluctuations on super-Hubble scales describes the growth of
cosmological perturbations, and this corresponds to particle production.
If the equation of state of the background cosmology is constant
in time, then the squeezing of the cosmological fluctuations occurs
at the same rate as the squeezing of test fields (which is the same
as the squeezing rate of gravitational waves).

In the case that the background cosmological space-time metric is 
constant in time both in the far past and in the future,
then we know that the wave function of a test scalar field can be written 
both in the far past and in the future as
\be \label{modeexp}
\chi(x, t) \, = \, \frac{1}{\sqrt{2 \pi}} V^{-1/2} \int d^3k 
\bigl[ \alpha_k \psi_k^+ + \beta_k \psi_k^-  \bigr]
\ee
in terms of coefficient functions $\alpha_k$ and $\beta_k$.
Here, $\chi_k^+$ and $\chi_k^-$ are the Minkowski
space-time positive and negative frequency wave
functions
\ba
\chi_k^+ \, &=& \, \frac{1}{\sqrt{2k}} e^{i kx} \nonumber \\
\chi_k^- \, &=& \, \frac{1}{\sqrt{2k}} e^{- i kx} \, .
\ea

If we start the evolution in the vacuum state $|\Psi>_i$ of the
system (the harmonic oscillator vacuum of each Fourier
mode), then initially $\alpha_k = 1$ and $\beta_k = 0$.
This is the Bunch-Davies \cite{BD} vacuum. An observer
at late times will define a new vacuum state $|\Psi>_f$ 
in which the coefficient functions have vanishing $\beta_k$
and $\alpha_k = 1$ at the final time. Because of
the time dependence of the background, the state
of the system $|\Psi>$ which equals the initial vacuum
at early times will evolve non-trivially and turn into
a state for which at the final time $\beta_k \neq 0$.
The late time coefficients $\alpha_k$ and $\beta_k$ are
called the Bogoliubov coefficients, and they obey the
relation (see e.g. \cite{BirD})
\be
\alpha^2_k - \beta^2_k \, = \, 1 \, .
\ee 
The interpretation of this state for the late time observer
is that of a state which contains
\be \label{number}
n_k \, = \, |\beta_k|^2 
\ee
particles of comoving wave number $k$. 

In most cosmological models (in particular in 
inflationary cosmology and in the Ekpyrotic scenario) 
the metric is neither static initially nor today. However,
modes which are relevant to current cosmological observations 
are inside the Hubble radius both at very early times and
today. On scales smaller than the Hubble radius the mode
functions of the canonically normalized fields \footnote{These
are fields with canonical kinetic term. For a test scalar field
$\chi$ the canonical field is $a(t) \chi$.} oscillate, and hence
the mode wavefunction can be represented in the form
(\ref{modeexp}). In the early universe scenarios which have
a chance of explaining the origin of the observed structure
in the Universe the modes exit the Hubble radius during
the initial phase (e.g. the inflationary phase in the case of
inflationary cosmology, or the Ekpyrotic phase of contraction
in the case of Ekpyrotic cosmology), and re-enter at late
times. While the scale is super-Hubble, we can still
expand the mode functions at any time $t$ in terms of
a local Minkowski frame in the form (\ref{modeexp})
with time-dependent Bogoliubov coefficients. However,
the wave function oscillations are frozen out, the
state is a squeezed state, adiabaticity is violated and
particle number is not a well-defined quantity (see e.g.
the discussion in \cite{Jerome2}). Hence the Bogoliubov
coefficients should not be interpreted as yielding the
number of particles. Heuristically, one could say that
via (\ref{number}) the Bogoliubov coefficients yield
the number of proto-particles, a field state which will
admit a particle interpretation once the scale enters
the Hubble radius.

\section{Parker Particle Production in the New Ekpyrotic Scenario}

The Ekpyrotic scenario \cite{Ekp} is an alternative to
cosmological inflation which was originally motivated
by some ideas in superstring theory, in particular 
heterotic M-theory \cite{heterotic}. In the Ekpyrotic scenario,
our universe emerges from an initial contracting phase which
arises when two three-space-dimensional branes (one of which
corresponds to our space-time) bounding
an extra spatial dimension approach each other. In the
original scenario, the branes collide and this corresponds
to a ``Big Crunch'' singularity after which our universe emerges
in a Standard Big Bang phase of expansion. 

The effective gravitational theory in our space-time is
Einstein gravity coupled to a scalar field $\phi$ (which
is proportional to the logarithm of the brane separation)
whose potential was assumed to be a steep negative exponential
\be
V(\phi) \, = \, - V_0 {\rm exp} \bigl( - \sqrt{\frac{2}{p}} \frac{\phi}{m_{pl}} \bigr) \, ,
\ee
with $0 < p \ll 1$ and $V_0 > 0$, and with $m_{pl}$
denoting the Planck mass. Inserting this into the Friedmann 
equations we find that the contraction is very slow
\be
a(t) \, \sim \, (-t)^{p} \, .
\ee
This corresponds to an equation of state with
\be
w \, \equiv \, \frac{p}{\rho} \, \gg 1 \, .
\ee
In turn, this equation of state implies that the energy
density $\rho_{\phi}$  stored in the field $\phi$ increases as
\be
\rho_{\phi} \, \sim \, a^{- 2 / p} \, ,
\ee
which implies that it increases faster than that in regular
cold matter, radiation, curvature and anisotropic stress.
In particular, in contrast to contracting phases with usual
matter content with $w = 0$ or $w = 1/3$, the contracting
phase is safe against the BKL instability \cite{BKL} of the 
homogeneous bounce, as shown explicitly in \cite{noBKL}.
This is a significant advantage of the Ekpyrotic scenario
compared to most other bouncing models (see e.g. \cite{Peter2}
for a discussion of the instability for regular bouncing models,
and the last entry of \cite{RHBrevs} for a recent review of 
problems of regular bouncing models). 

As a consequence of the slow contraction, fixed comoving
scales exit the Hubble radius during the period of contraction
since the Hubble radius decreases as $(-t)$. Hence, a
causal generation mechanism of fluctuations is possible.
Although the spectrum of $\phi$ fluctuations produced
during Ekpyrotic contraction is scale-invariant \cite{Ekp2},
that of curvature perturbations is not \cite{noscaleinv}. 

To solve this problem, the {\it New Ekpyrotic scenario}
was proposed \cite{new} (see also \cite{new2}). The model
involves two scalar fields $\phi$ and $\psi$, both with
negative exponential potentials
\be
V(\phi, \psi) \, = \, - V_0 {\rm exp} \bigl( - \sqrt{\frac{2}{p}} \frac{\phi}{m_{pl}} \bigr) 
- U_0 {\rm exp} \bigl( - \sqrt{\frac{2}{q}} \frac{\psi}{m_{pl}} \bigr) \,
\ee
with $p \ll 1$ and $q \ll 1$. Since there are two
fields present, it is possible to have entropy fluctuations. In
the same way that test scalar fields in an Ekpyrotic
background acquire a scale-invariant spectrum of fluctuations,
in the two field Ekpyrotic scenario the entropy mode
acquires a scale-invariant spectrum, and transmits this
spectrum to the curvature fluctuations since any entropy
fluctuation induces a growing curvature perturbation \footnote{This
is a standard result in the theory of cosmological perturbations.
For a recent study see e.g. \cite{Elisa}.}.

We are interested in a version of the New Ekpyrotic scenario
in which new physics (which involves, from the point of
view of Einstein gravity, a violation of the Null Energy Condition)
leads to a nonsingular bounce occurring when the background
density is 
\be
\rho_{max} \, = \, M^4 \, ,
\ee
where $M$ is the mass scale of the new physics leading to
the nonsingular bounce.

In the New Ekpyrotic scenario the background trajectory
is given by
\ba
a(t) \, &\sim& \, (-t)^{p + q} \, \\
\phi(t) \, &=& \,  \sqrt{2p} m_{pl} {\rm log} \bigl( - \sqrt{\frac{V_0}{m_{pl}^2 p (1 - 3(p + q))}}t \bigr) \, \nonumber \\
\psi(t) \, &=& \,  \sqrt{2q} m_{pl} {\rm log} \bigl( - \sqrt{\frac{U_0}{m_{pl}^2 q (1 - 3(p + q))}}t \bigr) \, \nonumber \, .
\ea
The field space is two-dimensional. Fluctuations in the field space direction parallel
to the background trajectory form the adiabatic mode $\sigma$ given by \cite{Gordon}
\be
{\dot \sigma} \, = \, {\rm cos} \theta {\dot \phi} + {\rm sin} \theta {\dot \psi} \,
\ee
and has adiabatic fluctuations
\be
\delta \sigma \, = \,  {\rm cos} \theta \delta \phi + {\rm sin} \theta \delta \psi
\ee
while the orthogonal direction $s$ has perturbations
\be
\delta s \, = \, - {\rm sin} \theta \delta \phi + {\rm cos} \theta \delta \psi \, .
\ee 

The canonically normalized entropy field perturbation is
\be
v \, \equiv \, a \delta s \, ,
\ee
and obeys the Fourier space equation
\be \label{veq1}
v'' + k^2 v - \frac{2}{\eta^2} \bigl(1 - \frac{3}{2}(p + q) \bigr) v \, = \, 0 \, ,
\ee
where $\eta$ is conformal time related to physical time via $dt = a d\eta$,
and where a prime denotes the derivative with respect to $\eta$. Note
that we are suppressing the index $k$ on the Fourier mode of $v$. In
the following we will study the production of $v$ particles in the
contracting phase of the Ekpyrotic scenario \footnote{Here, ``particle''
has to be interpreted in the sense described at the end of the previous
section.},

The initial conditions in Ekpyrotic cosmology are taken to be vacuum,
i.e. all fields begin in their Bunch-Davies vacuum state at past infinity.
All modes undergo quantum vacuum oscillations until their wavelength
crosses the Hubble radius, after which they will be squeezed. The
squeezing corresponds to particle production (as discussed in the
previous section). The modes on cosmological scales develop into
the density perturbations which we observe today, modes on microscopic
wavelengths (but larger than the Hubble radius at the end of the
contracting phase) become - at Hubble radius re-entry, when the
concept of particles becomes well-defined - the particles whose production we
are interested in. We are interested in computing the energy density in particles
produced by the end of the Ekpyrotic phase of contraction, {\it i.e.} when the
density reaches $\rho_{max}$.

We will consider a slightly more general setup which can be
applied not only to the New Ekpyrotic scenario, but also
to others. The equation of motion generalized from (\ref{veq1} is
\be \label{veq2}
v'' + \left(k^2-\frac{z''}{z}\right)v'' \, = \, 0 \,,
\ee
where $\eta$ is the conformal time which in a contracting background 
goes from $-\infty$ to $0$. We posit
\be
\frac{z''}{z} \, = \, \frac{\nu^2-1/4}{\eta^2} \, ,
\ee
with
\be
\nu \, \equiv \, \frac{1+{\tilde p}}{2(1- {\tilde p})} \, ,
\ee
where the value of ${\tilde p}$ depends on the specific model.
In the case of New Ekpyrotic scenario we have (see \ref{veq1})
\be
\nu \, = \, \sqrt{\frac{9}{4}-3(p+q)} \, .
\ee
There is an ``effective'' Hubble radius which divides modes
which oscillate from those which are squeezed. For a mode
with comoving wave number $k$, the conformal time $\eta_H(k)$
of effective Hubble radius crossing is given by
\be \label{Hubble}
k^2 \eta_H(k)^2 \, = \, \nu^2 - \frac{1}{4} \, .
\ee

At the beginning of the contraction phase all scales we are
interested in are inside the effective Hubble radius and hence the $k^2$ term  
dominates over the $z''/z$ term, and we start with the Bunch-Davies solution
\be
v \, = \, v_{BD} \, \equiv  \, \frac{e^{-ik\eta}}{\sqrt{2k}} \, .
\ee
After effective Hubble radius crossing (which occurs at the time
$\eta = \eta_H(k)$), the $k^2$ is subdominant then we have as 
solution  
\be \label{vsol}
v \, = \, c_1(k) \frac{\eta^{1/2-\nu}}{\eta_H(k)^{1/2-\nu}} 
+ c_2(k) \frac{\eta^{1/2+\nu}}{\eta_H(k)^{1/2+\nu}}
\ee
where $c_1(k)$ and $c_2(k)$ are constants which are found by matching 
$v$ and $v'$ at effective Hubble radius crossing $\eta = \eta_H(k)$. Then
\begin{eqnarray}
c_1(k) \, &=& \, \frac{-1}{\sqrt{2k}} e^{- i k \eta_H(k)} [\frac{1}{2} + \nu + i k \eta_H(k)] \, , \nonumber \\ 
c_2(k)  \, &=& \, \frac{1}{\sqrt{2k}} e^{- i k \eta_H(k)} [\frac{3}{2} + \nu + i k \eta_H(k)] \, .
\end{eqnarray}
The Bogoliubov coefficient $\beta_k$ at a time $\eta$ closer to the bounce
can be obtained by expanding the solution (\ref{vsol}) in terms of the local 
Bunch-Davies state given by $v_{BD}$ at the time $\eta$ (where $k |\eta| \ll 1$)
\begin{eqnarray}
v \, &=& \, \alpha_k v_{BD} + \beta_k v_{BD}^* \nonumber \\
v' \, &=& \, \alpha_k v'_{BD} + \beta_k v_{BD}'^* 
\end{eqnarray}
where the star stands for complex conjugation. Keeping only
the growing solution on super-Hubble scales this yields
\be \label{betaeq}
\beta(\eta) \, = \, \frac{c_1(k) \sqrt{2k}}{2} \bigl( \frac{\eta}{\eta_H(k)} \bigr)^{1/2 - \nu} 
[1 + \frac{1/2 - \nu}{i k \eta}] \, .
\ee

In case of the original Ekpyrotic model we have $\nu \approx 1/2$ and thus
\be
\beta_k \, = \, \frac{1}{2} e^{- i k \eta_H(k)} [1 + i k \eta_H(k) ] \, .
\ee
Hence
\be
n_k \, = \,  |\beta_k|^2 \,  = \,  \frac{1}{4} [1 + (k \eta_H(k))^2] \, ,
\ee
which for $k\eta_H(k) \ll 1$ is $n_k \approx 1/4$. Hence, there is no
significant particle production until the end of the contracting phase
in the original Ekpyrotic scenario.

In the case of New Ekpyrotic model $\nu \approx 3/2$. At first sight it
looks like the second term on the right hand side of (\ref{betaeq}) will
dominate. However, the contribution of this term does not have an
interpretation as particles \footnote{Another argument for neglecting
this term is the following: once the scales re-enter the Hubble radius
and the particle interpretation becomes valid, this term is small compared
to the one we are keeping.}. As mentioned at the end of the previous
section, the particle interpretation only applies when scales re-enter
the Hubble radius. At that time, the solution for $v$ should be
interpreted as a standing wave which decays equally into a right-moving
and left-moving wave. Hence, the value of $\beta$ is given by
half of the ratio of the final amplitude of $v$ to the initial amplitude,
and we have
\ba
\beta_k(\eta) \, &=& \, \frac{c_1(k) \sqrt{2k}}{2} \bigl( \frac{\eta}{\eta_H(k)} \bigr)^{-1}  \\
&=& e^{- i k \eta_H(k)} [1 + i k \eta_H(k)] \bigl( \frac{\eta}{\eta_H(k)} \bigr)^{-1} \, . \nonumber
\ea
The corresponding number density of produced particles is
\be \label{nodens}
n_k(\eta) \, = \, [1 + (k \eta_H(k))^2] \bigl( \frac{\eta}{\eta_H(k)} \bigr)^{-2} \, .
\ee

We now can compute the energy density $\rho_p(\eta)$ in the produced particles 
(strictly speaking it is the energy density which the state will have once the scales
re-enter the Hubble radius at late times) at the end of the Ekpyrotic phase
of contraction, a time we denote by $t_{end}$. We choose to normalize the
scale factor such that $a(t_{end}) = 1$, and thus physical and comoving momenta
coincide at this time. We have
\be
\rho_p(\eta) \, = \, \frac{1}{(2 \pi)^3} \int_0^{k_H(\eta_{end})} n_k \, k \, \, d^{3}k \, ,
\ee
where the final factor of $k$ represents the energy of the mode once it
starts to oscillate. Inserting (\ref{nodens}) and making use of the relation
(\ref{Hubble}) to determine $k_H(\eta)$ in terms of $\eta$, and in
addition using the fact that for Ekpyrotic contraction $\eta_{end} \simeq t_{end}$
we find
\be
\rho_p(t_{end}) \, \sim \, t_{end}^{-4} \, ,
\ee
which is to be compared with the background density $\rho_{bg}$
\be
\rho_{bg}(t_{end}) \, \sim \, t_{end}^{-2} m_{pl}^2 \, .
\ee
If the background density at the final time is given by $M^4$ in
terms of a ``new physics mass scale'' then we find
\be
\frac{\rho_p(t_{end})}{\rho_{bg}(t_{end})} \, \sim \, \bigl( \frac{M}{m_{pl}} \bigr)^4 \, .
\ee
This result implies that if the scale of new physics is high (e.g. between the
scale of particle physics ``Grand Unification'' and the Planck scale),
 then a sufficiently high density of particles is produced to lead to
post-bounce hot big bang phase beginning at temperatures not much
lower than that of Grand Unification. Note that the energy density which
at the bounce remains in the Ekpyrotic field rapidly redshifts relative
to that in the produced particles after the bounce.

\section{Conclusions and Discussion}

We have studied Parker particle production in the contracting
phase of the New Ekpyrotic scenario and have found that the process is 
sufficiently efficient to lead to a hot thermal expanding universe beginning at
temperatures not much lower than that of Grand Unification provided that
the new physics scale which yields the bounce is higher than that
of Grand Unification. Hence, the scenario does not require a
separate physics sector to generate reheating.

Note that the distribution of particles produced by the Parker process
does not have a thermal spectrum, as is the case after preheating
in inflationary cosmology. What the Parker process does (and similarly
preheating) is to convert energy from the homogeneous condensate
to particle quanta. These quanta can then interact and thermalize
by particle scattering. Since for Parker particle production the 
spectrum is peaked in the ultraviolet (as opposed to the spectrum 
after preheating which is suppressed in the ultraviolet) 

We have only considered Parker particle production in the
contracting phase of the Ekpyrotic scenario. This leads to
a lower bound on the total number of particles produced
during the entire cosmological evolution. The reason is twofold. 
First, particles will also be produced during the bounce phase and
in the post-bounce expanding phase, and this will add to the
total number of particles produced. However, the bounce phase
is expected to be short compared to a Hubble expansion time
at the end of the Ekpyrotic phase, and hence we do not
expect Parker particle production to be important during
the bounce. During the post-bounce phase of
radiation-dominated expansion there is no squeezing
of the fluctuations, and hence we do not expect Parker
particle production during most of the post-bounce
period. The second point which supports our statement
that we have computed a lower bound on the effectiveness
of Parker particle production is that the dominant mode will
continue to be squeezed on super-Hubble scales after
the Ekpyrotic phase of contraction ends, and hence the
particles produced during the contracting phase will remain.

Note that the method we used to analyze Parker particle
production is applicable to a wide range of models. The
application to the {\it matter bounce} scenario was already
considered in \cite{Jerome}. For all contracting models with 
$\nu \neq 1/2$ Parker particle production may be important.
 
\section*{Acknowledgement}
\noindent

One of us (RB) wishes to thank the Institute for Theoretical Studies of the ETH
Z\"urich for kind hospitality. RB acknowledges financial support from Dr. Max
R\"ossler, the Walter Haefner Foundation, the ETH Z\"urich Foundation, and
from a Simons Foundation fellowship. The research of RB is also supported in
part by funds from NSERC and the Canada Research Chair program. WSHR is grateful 
for the hospitality of the Physics Department of McGill University. 
WSHR was supported by Brazilian agency CAPES (proccess No 99999.007393/2014-08).
EF and LG acknowledge financial support from $CNP_q$ (Science Without Borders). 
We also wish to thank J. Quintin for useful discussions.

\end{document}